\documentclass[fleqn,review]{elsarticle}

\usepackage{lineno,hyperref}
\usepackage{amsmath, amssymb, mathtools}
\usepackage[normal]{subfigure}
\usepackage{caption}
\usepackage{tabularx}
\usepackage{booktabs}
\modulolinenumbers[5]

\journal{arXiv}

\begin{document}

\begin{frontmatter}

\title{Distinct criticality of phase and amplitude dynamics in the resting brain}

\author[1,2]{Robert Ton}
\author[2,3]{Gustavo Deco}
\author[4,5,6]{Morten L. Kringelbach}
\author[7,8]{Mark Woolrich}
\author[1]{Andreas Daffertshofer\corref{mycorrefauthor}}
\address[1]{MOVE Research Institute Amsterdam, VU University Amsterdam, Van der Boechorststraat 9, 1081BT Amsterdam, The Netherlands.}
\address[2]{Center for Brain and Cognition, Computational Neuroscience Group, Universitat Pompeu Fabra, Carrer Tanger 122-140, 08018 Barcelona, Spain.}
\address[3]{Instituci\'{o} Catalana de la Recerca i Estudis Avançats (ICREA), Universitat Pompeu Fabra, Carrer Tanger 122-140, 08018 Barcelona, Spain.}
\address[4]{University Department of Psychiatry, University of Oxford, Oxford, OX3 7JX, UK.}
\address[5]{Department of Clinical Medicine, Center of Functionally Integrative Neuroscience, Aarhus University, 8000 Aarhus C, Denmark.}
\address[6]{Department of Neurosurgery, John Radcliffe Hospital, Oxford, OX3 9DU, UK.}
\address[7]{Oxford Centre for Human Brain Activity, University of Oxford, Oxford, UK.}
\address[8]{Oxford Centre for Functional MRI of the Brain, Nuffield Department of Clinical Neuroscience, University of Oxford, Oxford, UK. }

\begin{abstract}
Converging research suggests that the resting brain operates at the cusp of dynamic instability signified by scale-free temporal correlations. We asked if the scaling properties of these correlations differ between amplitude and phase fluctuations, which may reflect different aspects of cortical functioning. Using source-reconstructed magneto-encephalographic signals, we found power-law scaling for the collective amplitude and for phase synchronization, both capturing whole-brain activity. The temporal changes of the amplitude comprise slow, persistent memory processes, whereas phase synchronization exhibits less temporally structured and more complex correlations, indicating a fast and flexible coding. This distinct temporal scaling supports the idea of different roles of amplitude and phase in cortical functioning.
\end{abstract}

\begin{keyword}
\texttt{Power laws, Criticality, DFA, Amplitude, Phase}
\end{keyword}

\end{frontmatter}

\newpage
\section{Introduction}
\noindent
It has been proposed that the brain is in or near a critical state | its dynamics may be positioned at the border between spatiotemporal order and disorder, reminiscent of non-equilibrium phase transitions in thermodynamic systems \cite{LinkenkaerHansenetal2001, Beggs&Plenz2003,Beggs2008, Beggs&Timme2012}. The concept of brain criticality is attractive because critical systems display optimal performance on several characteristics such as information transfer \cite{Greenfield&Lecar2001, Beggs&Plenz2003}, wide dynamic range \cite{Kinouchi&Copelli2006, Shewetal2009}, information capacity
\cite{Haken2006, Shewetal2011}, and long-term stability \cite{Chialvo2010, Shew&Plenz2013}. Criticality relates closely to self-organization \cite{Baketal1987, Chialvo2010, Shew&Plenz2013}, which is considered crucial to cortical functioning \cite{Christensenetal1998, Bornholdt&Roehl2003, Haken2006b, Rubinovetal2011, Drosteetal2012}.

A hallmark of critical behavior is the presence of power laws \cite{Stanley1971, Beggs&Timme2012}. Power laws symbolize scale-free behavior, adopting the same form on all time scales: they are self-similar. Consider the case of a scale-free auto-correlation function $AC$. The corresponding power law obeys the form $AC\left( s \cdot \tau \right) = s^{2H} \cdot AC \left( \tau \right)$, i.e. if time $\tau$ is rescaled to $s \cdot \tau$, then the shape of $AC$ is preserved and only rescaled by a factor $s^{2H}$. The scaling exponent $H$ is referred to as the Hurst exponent \cite{Hurst1951} and qualifies the underlying correlation structure: $H\!=\!0.5$ corresponds to an uncorrelated, random process whereas $H\!>\!0.5$ indicates persistent, long-range correlations.

There is accumulating evidence for the presence of power laws in brain activity \cite{Boonstraetal2013, He2014}. Neural spikes come in avalanches that display scale-free distributions \cite{Beggs&Plenz2003}. Spectral distributions of encephalographic signals have $1/f$-structures \cite{Freeman&VanDijk1987, Dehghanietal2010, Heetal2010, Franssonetal2013, Leietal2015} and their auto-correlation structures also show power-law behavior \cite{LinkenkaerHansenetal2001, LinkenkaerHansenetal2007}. In the spatial domain, scale-free distributions have been observed in functional as well as neuroanatomical connectivity patterns \cite{Eguiluzetal2005, Bullmore&Sporns2009, Fraimanetal2009}.

Previous work has focused only on spatially local measures of brain activity \cite{Freeman&VanDijk1987, LinkenkaerHansenetal2001, Beggs&Plenz2003, LinkenkaerHansenetal2007, Boonstraetal2013, He2014} or considered pairs of nodes in networks \cite{Kitzbichleretal2009, Botcharovaetal2014} rather than analyzing global brain activity. In complex systems, the global activity can be very informative about the generating dynamical structure \cite{Haken2006}, in particular when studying critical behavior \cite{Sethnaetal2001}. Here, we adopted these concepts to investigate criticality in the brain. By using source-reconstructed magneto-encephalographic (MEG) signals we sought to disambiguate between the scaling characteristics of amplitude and phase fluctuations because they may resemble different aspects of cortical functioning.

\section{Methods}
\subsection{MEG data \& outcome variables}
\noindent
Magneto-encephalographic (MEG) signals of ten subjects were recorded and sampled at 1 kHz in eyes-closed resting state for approximately five minutes. After down sampling to 250 Hz, signals were beamformed onto a ninety-node brain parcellation, yielding ninety time series $y_{k}\left( t \right)$ per subject. Data were previously published by Cabral and coworkers \cite{Cabraletal2014}.

Signals $y_{k}\left( t \right)$ were filtered with a second-order IIR-bandpass filter in the alpha band (8-12 Hz) and (upper) beta band (20-30 Hz). With the Hilbert transform we constructed the analytic
signal and defined phase $\phi_{k}\left( t,f \right)$ and amplitude $a_{k}\left( t,f \right)$ as functions of time $t$; $f$ indexes either the alpha or the beta frequency band.

We used two collective variables to capture whole-brain activity per subject. First, we defined the phase synchronization,
\begin{align*}
R \left( t,f \right) = \frac{1}{90}\left \vert \sum_{k = 1}^{90}e^{i\phi_{k}\left( t,f \right)} \right \rvert
\end{align*}
and, second, the mean amplitude,
\begin{align*}
A \left( t,f \right) =  \frac{1}{90}\sum_{k = 1}^{90} a_{k} \left( t,f \right)
\end{align*}
We note that $R \left( t,f \right)$ is the modulo of the complex-valued Kuramoto order parameter \cite{Kuramoto1984}. We {\it z}-scored $R \left( t,f \right) $ and $A\left( t,f \right)$ to reduce
between-subject variability such that we could assess subject-averaged behavior by means of a detrended fluctuation analysis (DFA, \cite{Pengetal1994}), as described below.

\begin{figure}[!ht]
\centering 
\includegraphics[width=0.95\textwidth]{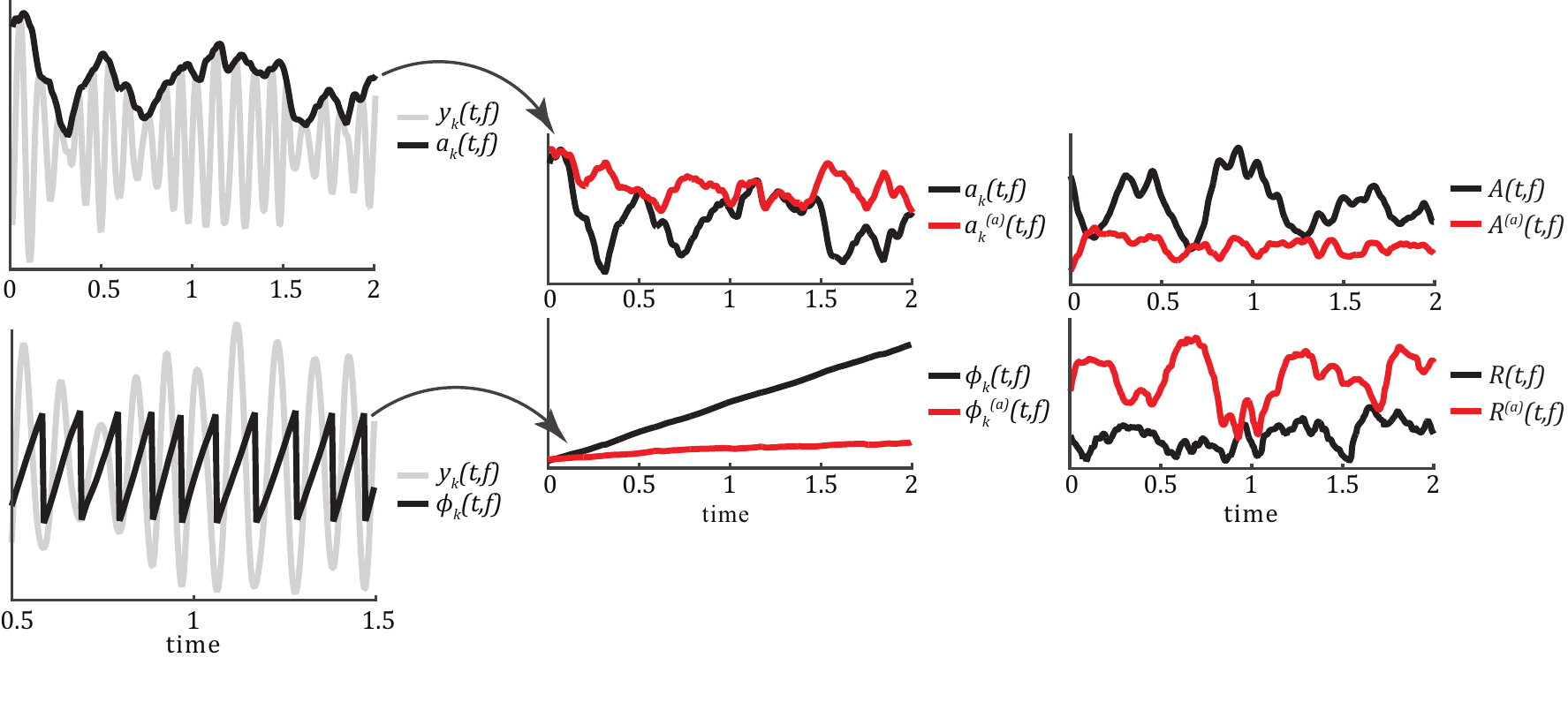}
\caption{\small {\bf Figure \ref{fig4:fig1}:} Signal $y_{k}\left( t,f \right)$ after filtering in the alpha band and corresponding Hilbert-amplitude $a_{k}\left( t,f \right)$ (upper left) and Hilbert-phase $\phi_k(t,f)$ (lower left). For clarity we decreased the time axis range and rescaled the $\phi_{k}\left( t,f \right)$ trace in the lower left panel. The amplitude is also displayed (black) in the upper middle panel together with its envelope $a_{k}^{(a)}\left( t,f \right)$ (red; see main text for the detailed definition). The Hilbert-phase $\phi_{k}\left( t,f \right)$, equal to $\phi_{k}\left( t,f \right)$ in the lower left panel, and the Hilbert-phase of $a_{k}\left( t,f \right)$, $\phi_{k}^{(a)}\left( t,f \right)$, are displayed in the lower middle panel. Different slopes indicate separate time scales (frequencies). The upper right panel shows amplitudes $A\left( t,f \right)$ and $A^{(a)}\left( t,f \right)$, the lower right one the phase order parameters $R\left( t,f \right)$ and $R^{(a)}\left( t,f \right)$}
\label{fig4:fig1}
\end{figure}

To relate our study to the previously established results in RSNs, we also examined the amplitude dynamics in more detail (see also \cite{Brookesetal2011}). The expression of RSNs is mainly reflected in the low-frequency content of the amplitudes $a_{k}\left( t,f \right)$, whose time scale is comparable to those of the blood-oxygenation-level-dependent (BOLD) signal \cite{Biswaletal1995}. In order to study these slow amplitude dynamics, we evaluated the longer time scales of the $a_{k}\left( t,f \right)$ dynamics by extracting its amplitude $a_{k}^{\left( a \right)}\left( t,f \right)$ and phase $\phi_{k}^{(a)}\left( t,f \right)$. Subsequently, we defined the collective variables associated with the amplitude dynamics:
\begin{align*}
R^{\left( a \right)}\left( t,f \right) &= \frac{1}{90} \left \lvert \sum_{k = 1}^{90} e^{ i\phi_{k}^{\left( a \right)}\left( t,f \right)} \right \rvert \\
A^{\left( a \right)}\left( t,f \right) &= \frac{1}{90} \sum_{k = 1}^{90} a_{k}^{\left( a \right) \left( t,f \right)}
\end{align*}
Further analysis was identical to that described above; see {\bf Figure \ref{fig4:fig1}} for illustration.\\
\hfill

\subsection{Data analysis | DFA}
DFA is considered robust against non-stationarity rendering it suitable for analyzing the temporal autocorrelation structure of encephalographic activity, in general, and global amplitude and phase synchronization, in particular. We employed a modified form of DFA including (Bayesian) model selection to verify the presence of power-law behavior \cite{Ton&Daffertshofer2015}.

In a nutshell, to quantify the autocorrelation structure of (the cumulative sum of) a signal $Y\left( t \right)$, one divides it into non-overlapping segments $Y_{i}\left( t \right)$, with $t = \ 1,...,n$ being discrete time steps and $i = 1,...,M$ indexing the segments; $M = \left\lfloor N/n \right\rfloor$ is the number of non-overlapping segments of length $n$. In each segment the linear trend $Y_{i}^{\text{trend}}\left( t \right)$ is removed providing an estimate of fluctuations in terms of
\begin{align*}
F_{i}\left( n \right)  =  \sqrt{\frac{1}{n}\sum_{t = 1}^{n} \left( Y_{i}\left( t \right) - Y_{i}^{\text{trend}}\left( t \right) \right)^{2}}
\end{align*}
This definition yields a set of `realizations' of fluctuations $F_{i}$ that, in the presence of a power law, scale like $F_{i}\left( n \cdot \tau \right) = \ n^{\alpha} \cdot F_{i}\left( \tau \right)$,
which is equivalent to $\log\left( F_{i} \right) =  \alpha \cdot \log \left( n \right) + \text{{\it const}}$. That is, in a log-log representation these fluctuations have a linear relationship with segment size. DFA seeks to identify the scaling exponent $\alpha$ that provides an estimate for the aforementioned Hurst exponent $H$.

\begin{figure}[!ht]
\centering 
\subfigure[]{\label{subfig4:fig2a}\includegraphics[width=0.45\textwidth]{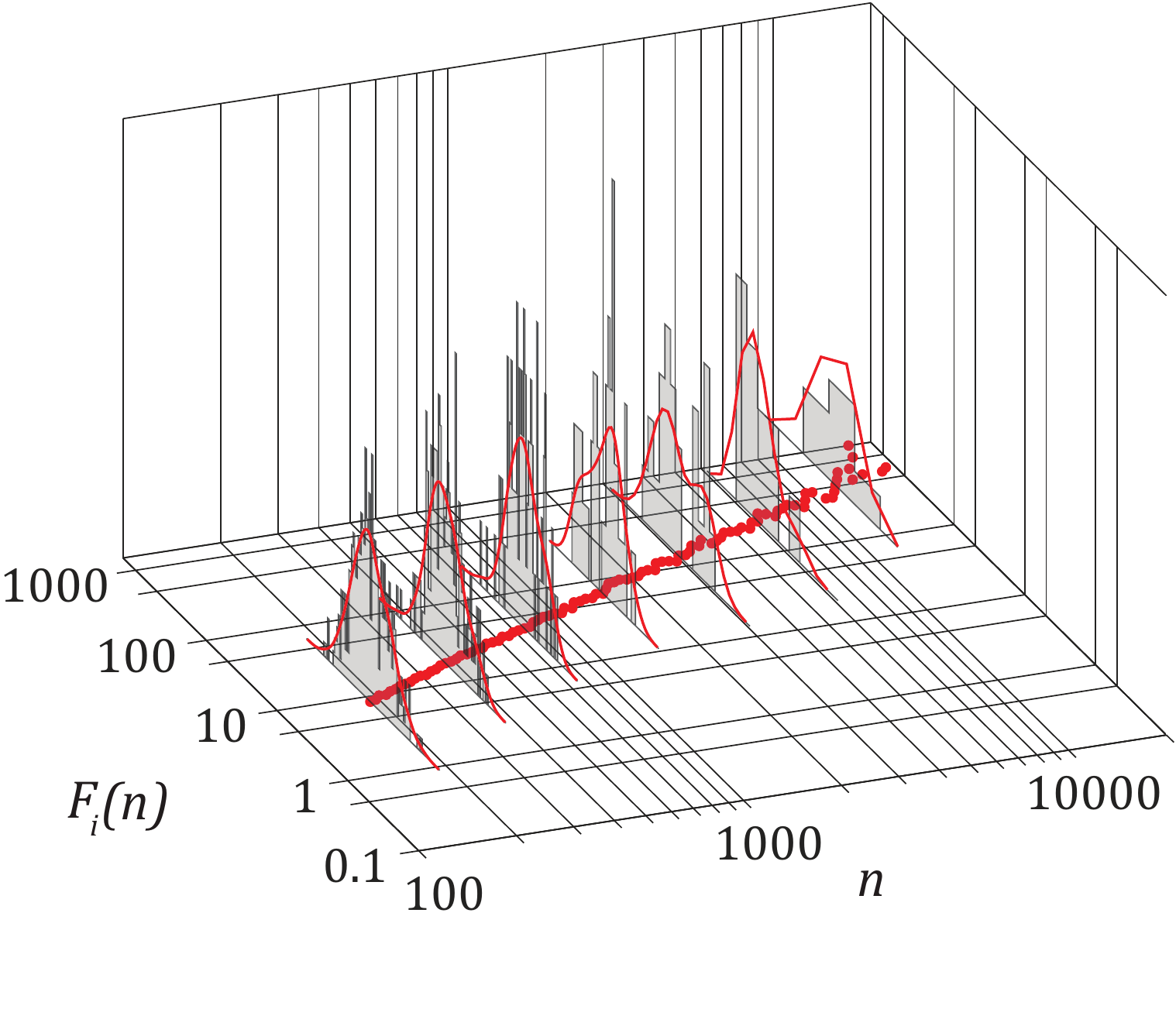}}
\subfigure[]{\label{subfig4:fig2b}\includegraphics[width=0.45\textwidth]{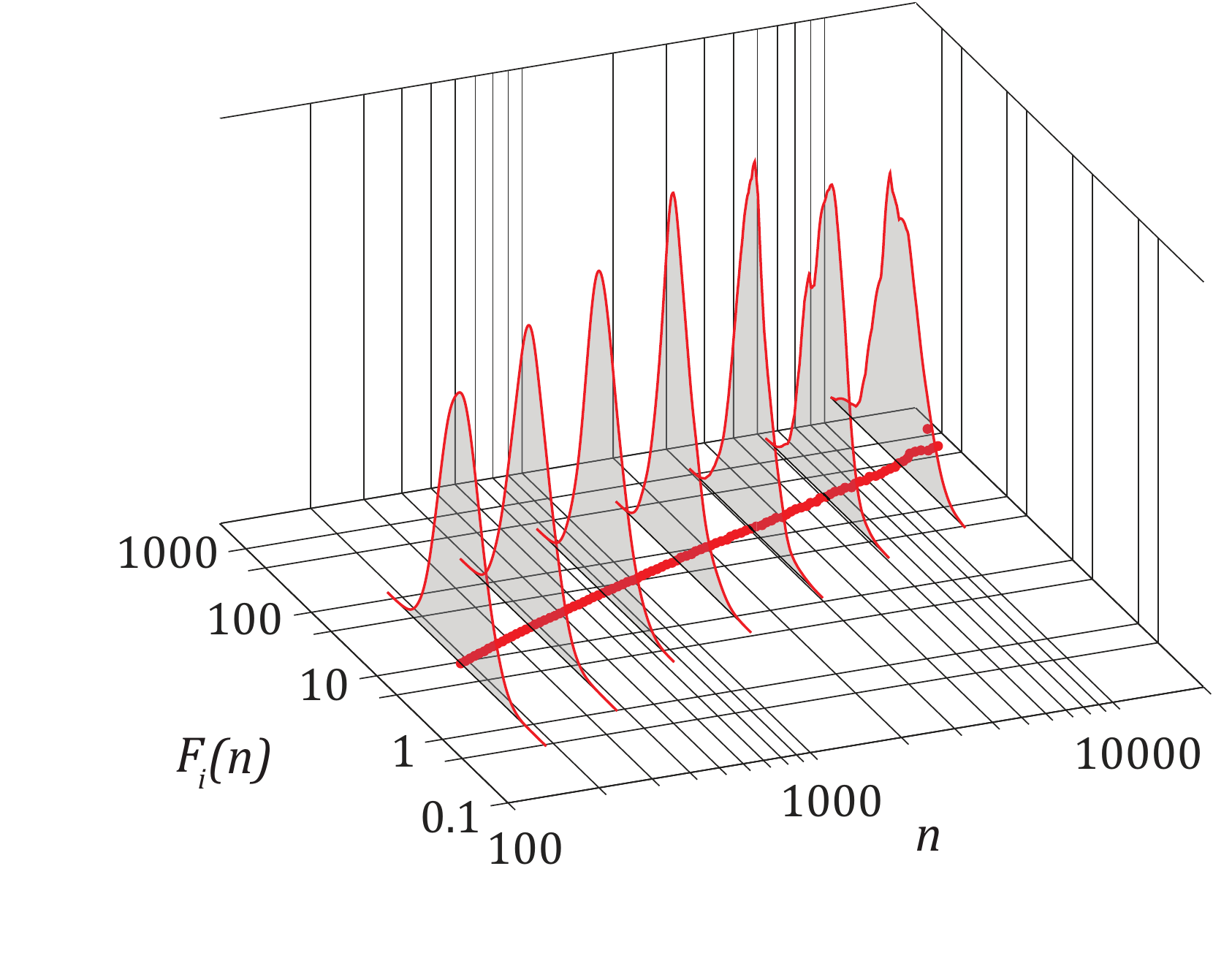}}
\caption{\small {\bf Figure \ref{fig4:fig2}:} Figure illustrating the relation between $F$ (red dots at the bottom) and the individual densities $p_{n}$ (red lines, {\bf Figure \ref{subfig4:fig2a}}) and averaged densities ${{\overline{}}{p}}_{n}$ (red lines, {\bf Figure \ref{subfig4:fig2b}}) or $A\left( t \right)$ in the alpha band; we show $p_{n}$ only for a few values of $n$. {\bf Figure \ref{subfig4:fig2a}} depicts the histograms of $F_{i}\left( n \right)\ $on basis of which $p_{n}$ were determined by kernel density estimation. The $F$ values are the expectation values of $p_{n}$ and ${\overline{p}}_{n}$, respectively.
}
\label{fig4:fig2}
\end{figure}

Instead of computing the mean value of $F_{i}$ as in conventional DFA \cite{Pengetal1994}, we here determined the probability density function $p_{n}\left( F_{i} \right)$ for every segment length $n$ (see {\bf Figure \ref{fig4:fig2}}). This approach allows for quantifying the appropriateness of a model $f_{\theta}\left( \tilde{n} \right)$ for fitting the fluctuation structure $\tilde{F} = \log\left( F_{i} \right)$ as a function of
$\tilde{n} = \log\left( n \right)$ by means of the log-likelihood function $\ln\left( \mathcal{L} \right)  =  \sum_{n}{\ln\left( \tilde{p}_{n}\left( f_{\theta} \right) \right)}$, where the tilde indicates a transformation to logarithmic coordinates. Here, the model $f_{\theta}\left( \tilde{n} \right)$, parametrized by the set $\theta$, may obey any arbitrary form including the linear one, which corresponds to a power law. We tested this linear relationship against a set of alternative models ({\bf Table \ref{tab4:tab1}}) using both the Bayesian information criterion and the Akaike information criterion (BIC and AIC$_c$, respectively). The model resulting in the least value of the information criterion was selected as the proper model.
Whenever this yielded the linear model $f_{\theta}^{1}$, we considered power-law behavior to be present and identified the scaling exponent with its slope \cite{Ton&Daffertshofer2015}. \\
\hfill
\begin{table}[h!]
\centering
\caption{\small {\bf Table \ref{tab4:tab1}:} The set of candidate models for the selection procedure. The linear model $f_{\theta}^{1}\left( x \right)$ is the form a power law would adopt. The alternative models $f_{\theta}^{2}$ - $f_{\theta}^{7}$ constitute polynomials up to third order. With $\ f_{\theta}^{8}\left( x \right)$ and $f_{\theta}^{9}\left( x \right)$ we considered two models resembling a (un)stable linear stochastic dynamics. } \label{tab4:tab1}
\begin{tabularx}{\textwidth}{p{0.4\linewidth}  l}
\toprule
$f_{\theta}^{1}\left( x \right)  = \theta_{1} + \theta_{2}x$ & $f_{\theta}^6\left( x \right)  = \theta_1 + \theta_2 x^2 + \theta_3 x^3$ \\
$f_{\theta}^2\left( x \right) = \theta_1 + \theta_2 x^2$ & $f_{\theta}^7 \left( x \right)  = \theta_1 + \theta_2 x + \theta_3 x^2 + \theta_4 x^3$ \\
$f_{\theta}^3 \left( x \right)  = \theta_1 + \theta_2 x + \theta_3 x^2$ & $f_{\theta}^8(x) = \theta_1 + \theta_2 e^{\theta_3 x}$ \\
$f_{\theta}^4 \left( x \right)  = \theta_1 + \theta_2 x^3$ & $f_{\theta}^9(x) = \theta_1 + \frac{1}{\ln(10)}  \ln \left( \theta_2 \left( 1 - e^{- \theta_3 e^{\ln(10) x}} \right) \right)$ \\
$f_{\theta}^{5}\left( x \right)\  =  \theta_1 + \theta_2 x + \theta_3 x^3 $ & \\
\bottomrule
\end{tabularx}
\end{table}

Since we were interested in the subject-averaged scaling exponents, we determined $p_{n}$ for every subject individually and averaged over subjects to obtain $\overline{p}_{n}$. These averaged probability density functions were used both for model selection and to determine the scaling exponent $\alpha$. The scaling range was given by
$\left[ n_{\min},~n_{\max} \right]  = \left[ 1.875 \cdot 10^{2},~1.875 \cdot 10^{4} \right]  \simeq \left[ 0.75,\ 75 \right]$ seconds, i.e. it spanned two decades. In this range we used one hundred equally spaced window sizes (on a logarithmic scale).\\
\hfill
\subsection{Statistics | surrogate data}
\noindent
Bootstrapping served to establish statistical significance using three types of surrogate data. Two of them consisted of randomly permuting temporal order whereas the third one only influenced cross-correlation structure. Of all types we constructed 1000 surrogates and significance values were obtained using $p  =  1  - \int_{- \infty}^{\alpha} p_{\text{surr}} (h)~dh$, where $p_{\text{surr}}\left( h \right)$ denotes the surrogate distribution and $\alpha$ the obtained empirical value of the scaling exponent.

For the first type of surrogates, we randomly permuted the order parameter time series for each subject. With this we evaluated our DFA and fitting procedure, since the surrogate time series lacked any temporal correlation structure and therefore should result in $\alpha = 0.5$ \cite{Mandelbrot&VanNess1968}. With the second type of surrogate we evaluated whether the filtering procedure and Hilbert analysis could have biased the results. For this, we permuted all original time series $y_k \left( t \right)$. By permuting
$y_k \left( t \right)$, all temporal structure was destroyed and therefore this constituted a rather weak null. In the third type of surrogates we performed a random cycling of $y_k\left( t \right)$ by shifting the time indices of $y_k \left( t \right)$ for each $k$, but keeping their order intact. In this way we retained the original auto-correlation structure of $\phi_k^{( \cdot )}(t,f)$ and $a_k^{( \cdot )}(t,f)$ but destroyed the cross-correlations. Subsequent analyses for all collective variables were identical to those for the original data.
\section{Results}
\begin{figure}[!ht]
\centering 
\subfigure[]{\label{subfig4:fig3a}\includegraphics[width=0.45\textwidth]{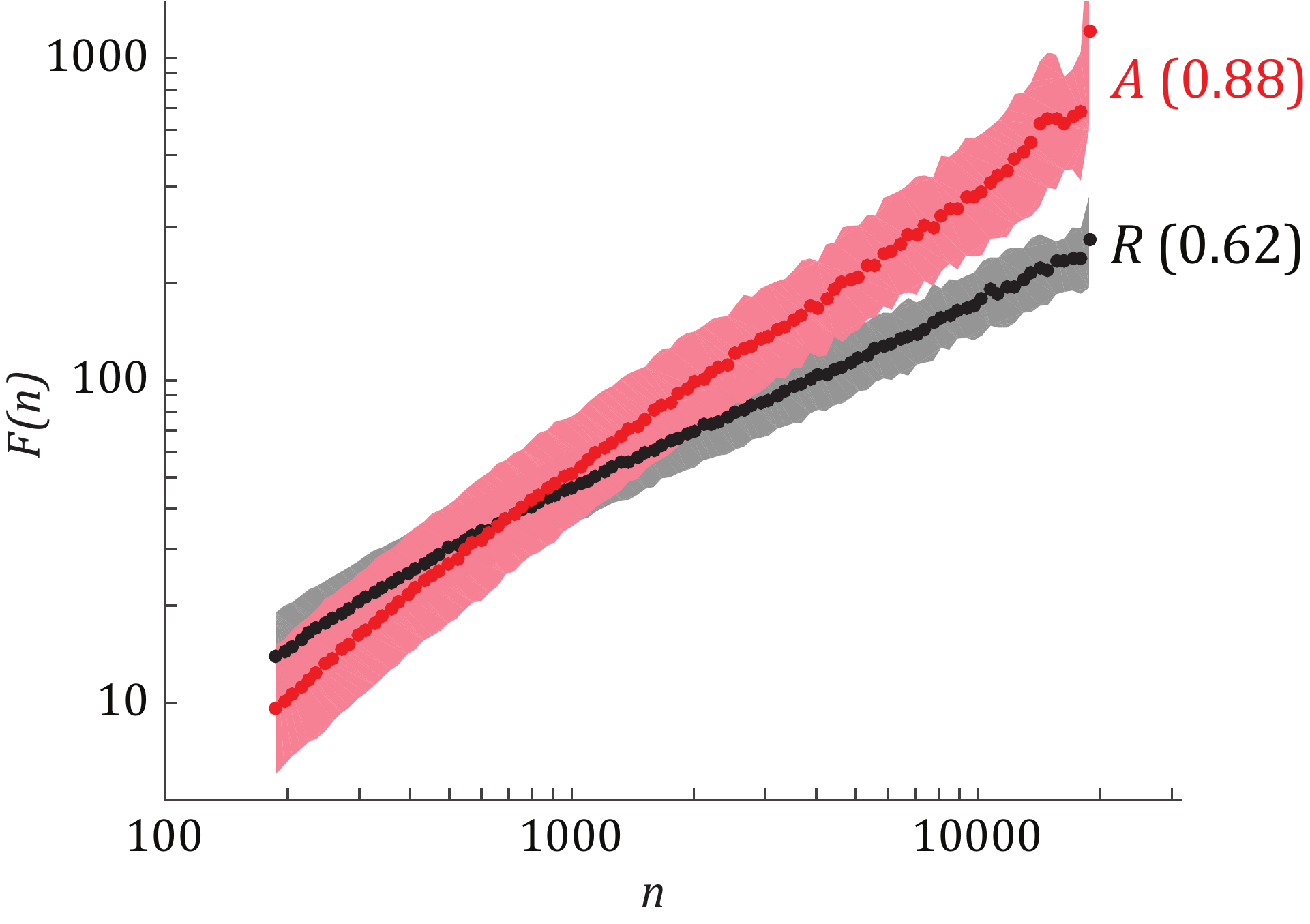}}
\subfigure[]{\label{subfig4:fig3b}\includegraphics[width=0.45\textwidth]{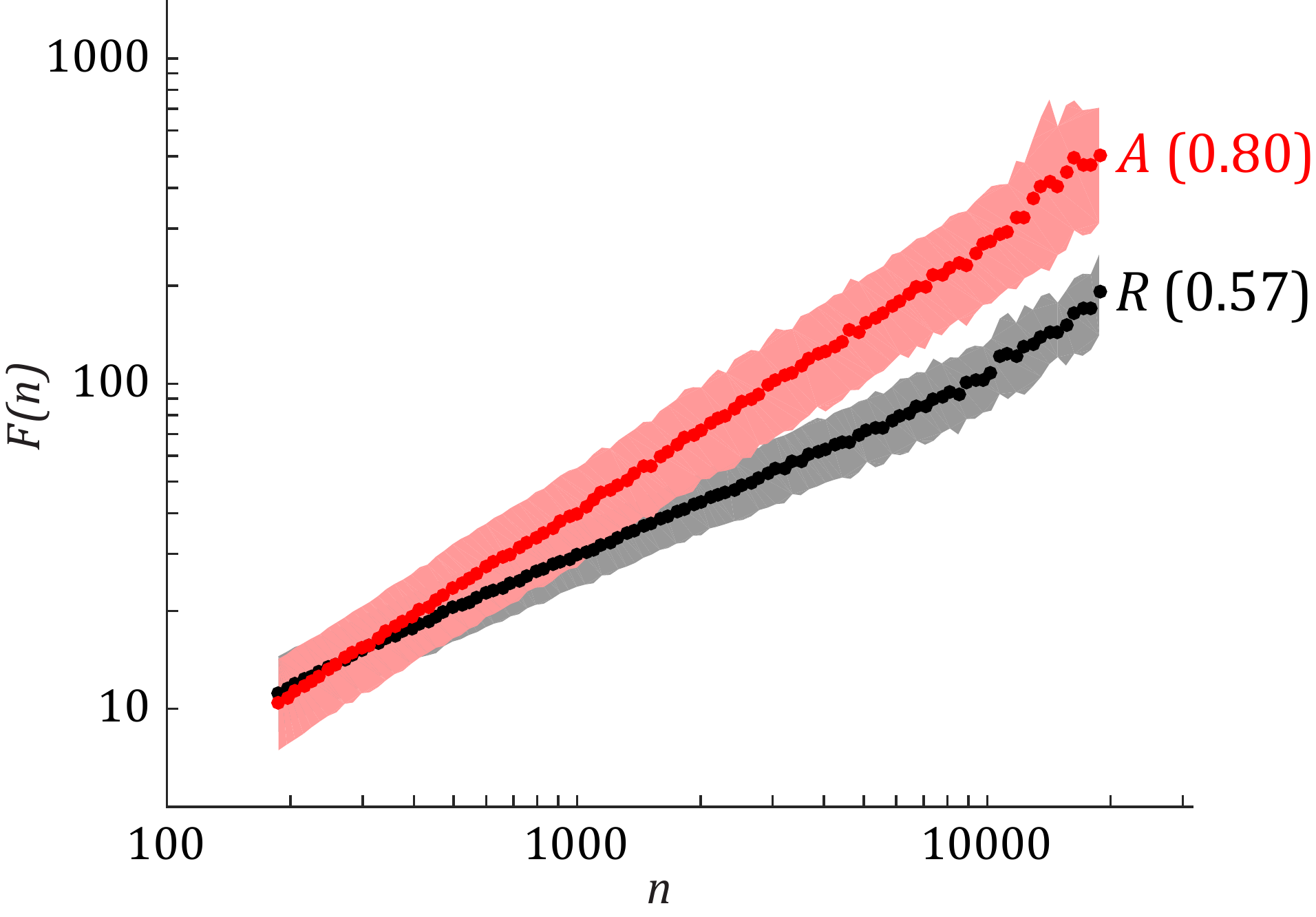}}
\caption{\small {\bf Figure \ref{fig4:fig3}:} Scaling behavior for $A\left( t \right)$ (red) and $R\left( t \right)$ (black). {\bf Figure \ref{subfig4:fig3a}} shows results for the alpha frequency band, {\bf Figure \ref{subfig4:fig3b}} for the beta frequency band. Scaling exponents are equal to 0.88 ($A$) and 0.62 ($R$) in the alpha and and 0.80 ($A$) and 0.57 ($R$) in the beta band. As in {\bf Figure \ref{fig4:fig2}} the dots display the expectation values of the subject-averaged probability densities with {\it F} on the vertical axis as function of window size $n$ on the horizontal axis. Shaded areas refer to the $25^{\text{th}}$ and $75^{\text{th}}$ percentiles of the subject averaged densities $\overline{p}_{n}$; see {\bf Tables \ref{tab4:tab2}-\ref{tab4:tab3}} for the model selection results. }
\label{fig4:fig3}
\end{figure}

\noindent
The presence of power-law scaling was evidenced by clear linear relationships of $A\left( t,f \right)$ and $R\left( t,f \right)$ fluctuations in log-log scale; see {\bf Figure \ref{fig4:fig3}}. This was confirmed by the AIC$_c$ and BIC values preferring the linear model in all cases ({\bf Tables \ref{tab4:tab2}-\ref{tab4:tab3}}). In both chosen frequency bands, the brain's network dynamics, as measured by the collective variables $A\left( t,f \right)$ and $R\left( t,f \right)$, thus appeared to exhibit scale-free correlations over a very broad range of time scales. \\
\hfill 

In the alpha band the scaling exponents were 0.88 and 0.62, and in the beta band 0.80 and 0.57, for amplitude $A$ and phase synchronization $R$, respectively. That is, both $A\left( t,f \right)$ and $R\left( t,f \right)$ showed persistent behavior but amplitude had increased perseverance compared to phase ({\bf Figure \ref{fig4:fig3}}). 

\begin{figure}[!ht]
\centering 
\subfigure[]{\label{subfig4:fig4a}\includegraphics[width=0.45\textwidth]{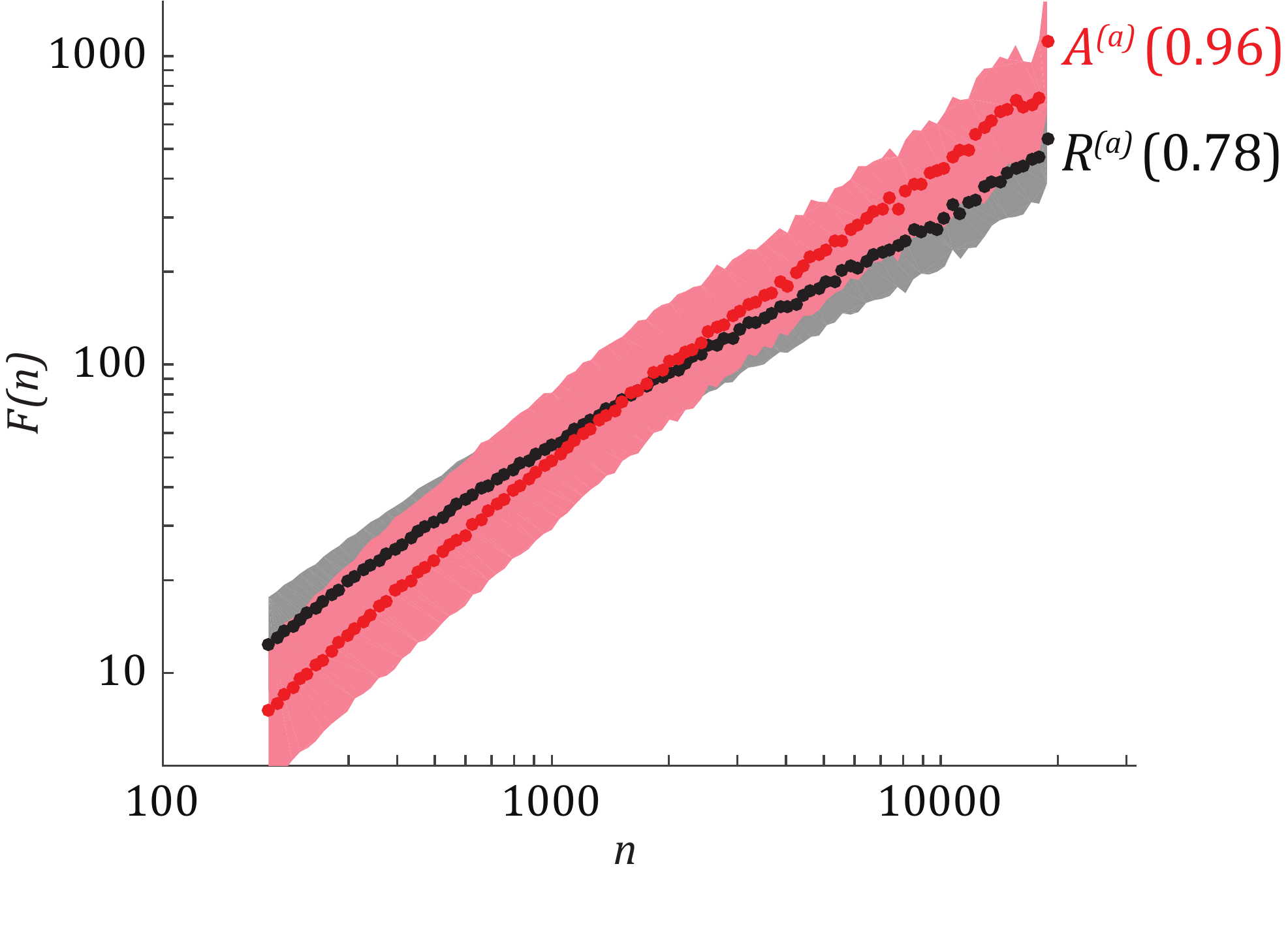}}
\subfigure[]{\label{subfig4:fig4b}\includegraphics[width=0.45\textwidth]{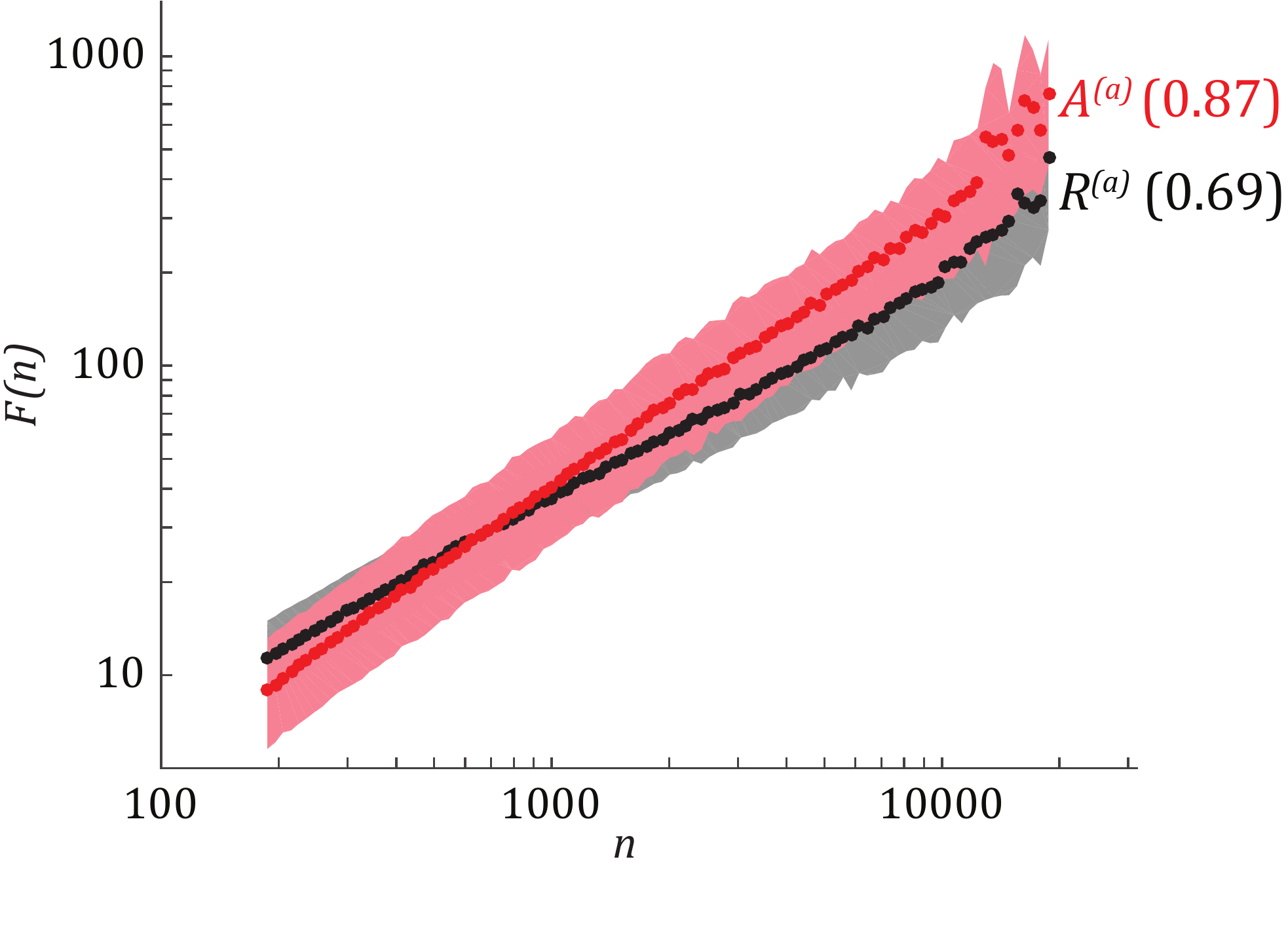}}
\caption{\small {\bf Figure \ref{fig4:fig4}:} Scaling behavior for $A^{\left( a \right)}\left( t \right)$ (red) and $R^{\left( a \right)}\left( t \right)$ (black). {\bf Figures \ref{subfig4:fig4a}, \ref{subfig4:fig4b}} show results for the alpha/beta frequency band. Scaling exponents $\alpha$ are equal to 0.96 ($A^{\left( a \right)}$) and 0.78 ($R^{\left( a \right)}$) in the alpha band and 0.87 ($A^{\left( a \right)}$) and 0.69 ($R^{\left( a \right)}$) in the beta band. Shaded areas refer to the $25^{\text{th}}$ and $75^{\text{th}}$ percentiles of the subject averaged densities $\overline{p}_n$; see {\bf Tables \ref{tab4:tab2}-\ref{tab4:tab3}} for the model selection results. }
\label{fig4:fig4}
\end{figure}

\begin{figure}[ht]
\centering 
\subfigure[]{\label{subfig4:fig5a}\includegraphics[width=0.45\textwidth]{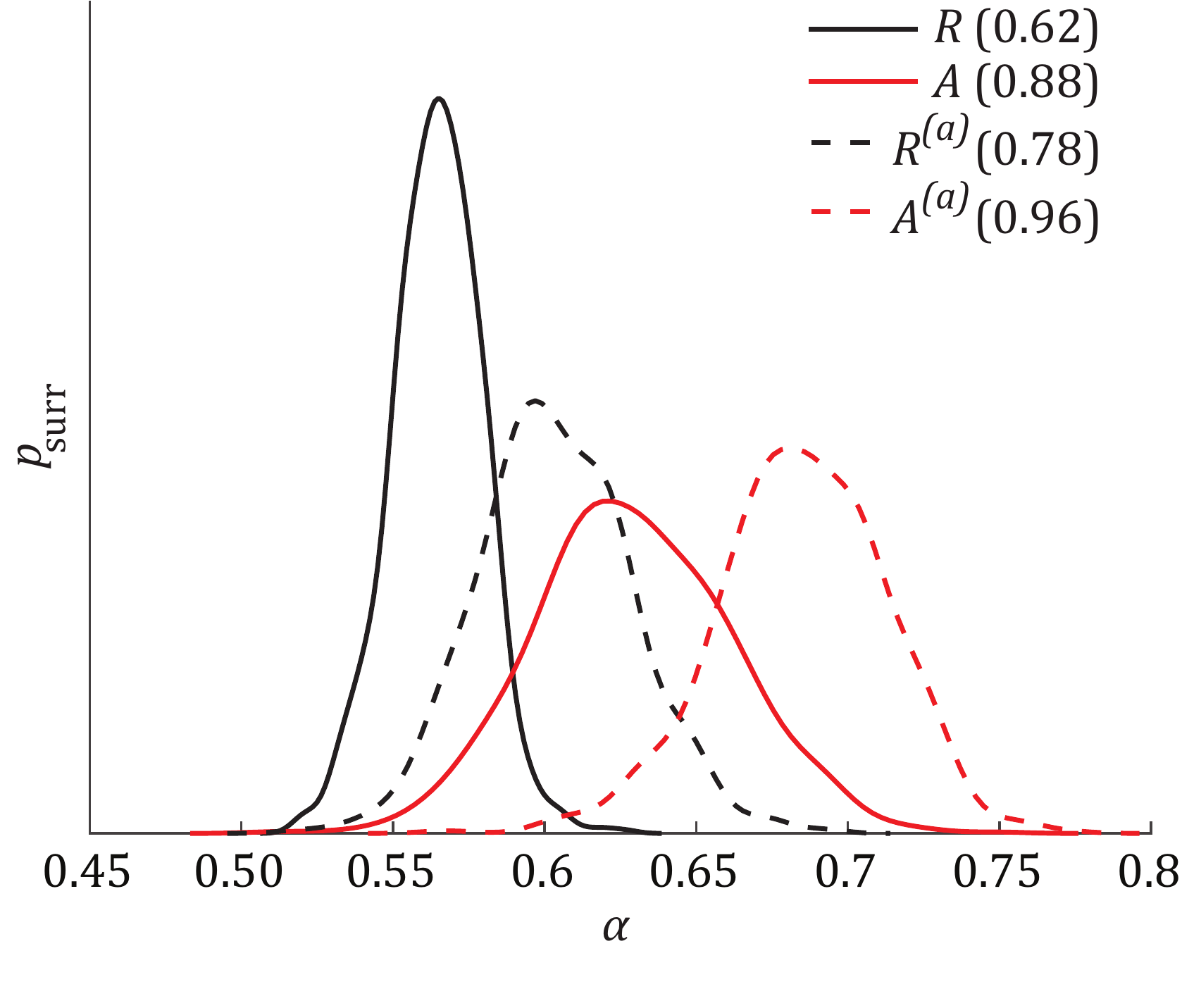}}
\subfigure[]{\label{subfig4:fig5b}\includegraphics[width=0.45\textwidth]{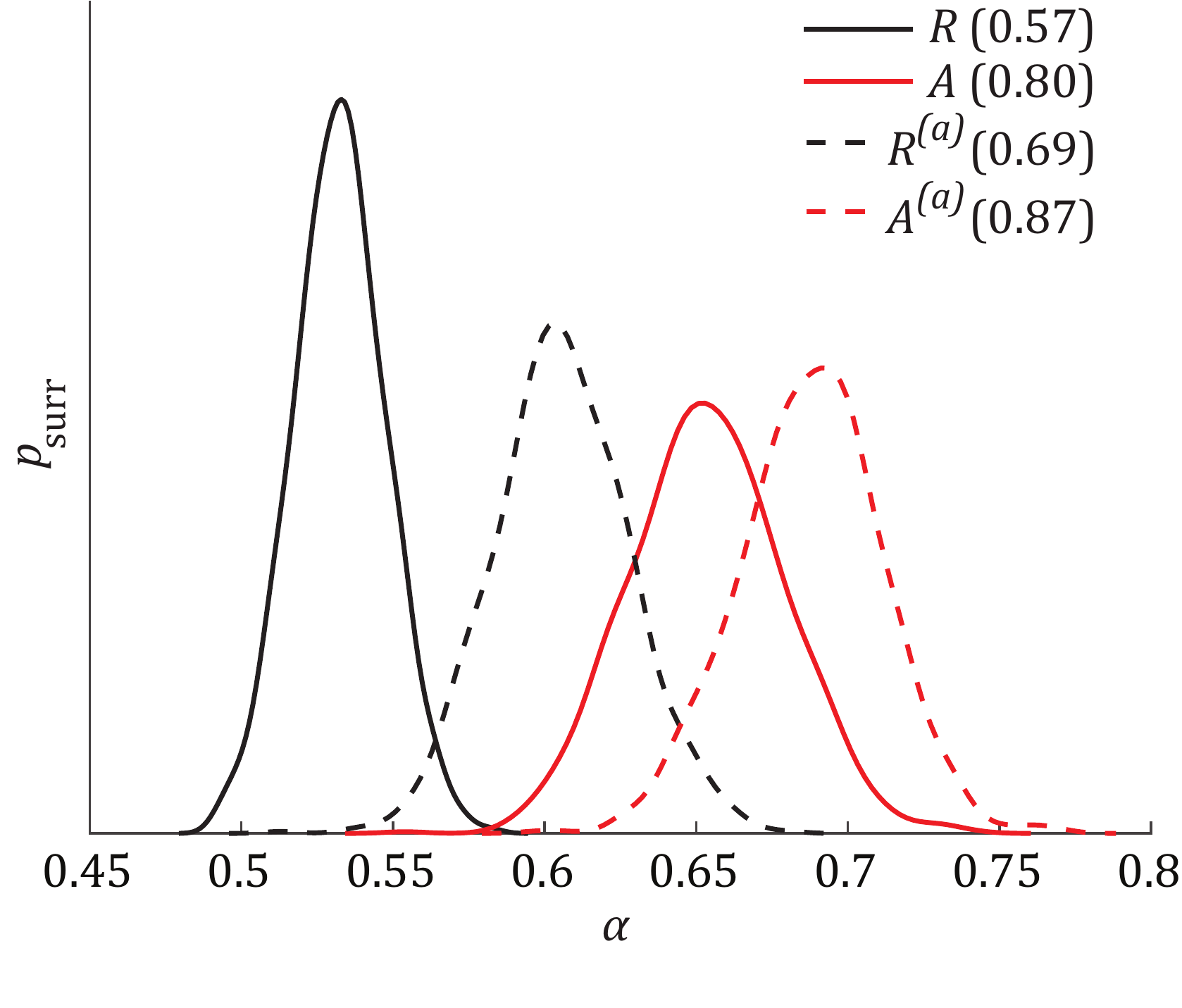}}
\caption{\small {\bf Figure \ref{fig4:fig5}:} Distributions of $\alpha$ values on basis of cycled time series $y_{k}\left( t \right)$ for the alpha band ({\bf Figure \ref{subfig4:fig5a}}) and the beta band ({\bf Figure \ref{subfig4:fig5b}}), obtained by applying a kernel smoothing method on the histograms for the order parameters $R\left( t,f \right)$ (black solid), $A\left( t,f \right)$ (red solid), $R^{(a)}\left( t,f \right)$ (black dashed) and $A^{(a)}\left( t,f \right)$ (red dashed). For reference scaling exponents are given in the legends. All original time series $\alpha$ values were significantly higher ($p < .01$) than those obtained from the surrogates. }
\label{fig4:fig5}
\end{figure}

We also found long-range temporal correlations in the variables $A^{\left( a \right)}\left( t,f \right)$ and $R^{\left( a \right)}\left( t,f \right)$ as shown in {\bf Figure \ref{fig4:fig4}} and confirmed by the AIC$_c$ and BIC values by preferring the linear model, except in case of the AIC$_c$ for $A^{(a)}(t,f)$ in the alpha band ({\bf Tables \ref{tab4:tab2}-\ref{tab4:tab3}}). The difference in scaling behavior between
$A^{\left( a \right)}\left( t,f \right)$ and $R^{\left( a \right)}\left( t,f \right)$ was similar to that between $A\left( t,f \right)$ and $R\left( t,f \right)$. \\
\begin{table}[htb]
\centering
\caption{ \small {\bf Table \ref{tab4:tab2}:} Model selection results for the alpha frequency band using $\text{BIC}  =  - 2 \ln \left( \mathcal{L} \right)  + K \ln \left( M \right)$
and $\text{AIC}_c  = - 2 \ln \left( \mathcal{L} \right)  +  2K  + \frac{\left( 2K \left( K - 1 \right) \right)}{\left( M - K - 1 \right)}$; with $K$ being the number of the parameters per model. The table shows relative values $\Delta \text{BIC}  = \text{BIC} - \min(\text{BIC})$. In all cases the linear model $f_{\theta}^{1}$ resulted in minimal BIC values indicating power-law scaling in all variables. Corresponding $\Delta \text{AIC}_{c}$ values are given in brackets.} \label{tab4:tab2}
\begin{tabularx}{\textwidth}{p{0.2\textwidth} r r r r}
\toprule
$f_{\theta}^{1}\left( x \right)$ & 0.00 (0.00) & 0.00 (0.00) & 0.00 (0.00) & 0.00 (0.10) \\
$f_{\theta}^{2}\left( x \right)$ & 5.85 (5.85) & 3.56 (3.56) & 4.92 (4.92) & 7.17 (7.27) \\
$f_{\theta}^{3}\left( x \right)$ & 4.60 (2.12) & 4.60 (2.12) & 4.60 (2.12) & 2.38 (0.00) \\
$f_{\theta}^{4}\left( x \right)$ & 16.21 (16.21) & 12.09 (12.09) & 16.95 (16.95)	 & 17.60 (17.70) \\
$f_{\theta}^{5}\left( x \right)$ & 4.60 (2.12) & 4.50 (2.02) & 4.58 (2.10) & 2.48 (0.10) \\
$f_{\theta}^{6}\left( x \right)$ & 4.89 (2.41) & 4.94 (2.46) & 5.08 (2.60) & 2.88 (0.50) \\
$f_{\theta}^{7}\left( x \right)$ & 9.02 (4.11) & 7.01 (2.10) & 8.29 (3.38) & 6.63 (1.82) \\
$f_{\theta}^{8}\left( x \right)$ & 4.60 (2.12) & 4.68 (2.20) & 4.60 (2.12) & 4.67 (2.30) \\
$f_{\theta}^{9}\left( x \right)$ & 32.05 (29.57) & 6.01 (3.53) &13.98 (11.50) & 3.69 (1.32) \\
\bottomrule
\end{tabularx}
\end{table}
\newpage
Testing against surrogate data confirmed the significance of these correlations. {\bf Figure \ref{fig4:fig5}} depicts the scaling exponent distributions corresponding to the third type of surrogates. For all variables in both frequency bands the scaling exponents significantly exceeded those of the surrogates ($p < .01$). \\
\hfill

\begin{figure}[ht]
\begin{minipage}[c]{0.45\linewidth}
\includegraphics[width = \linewidth]{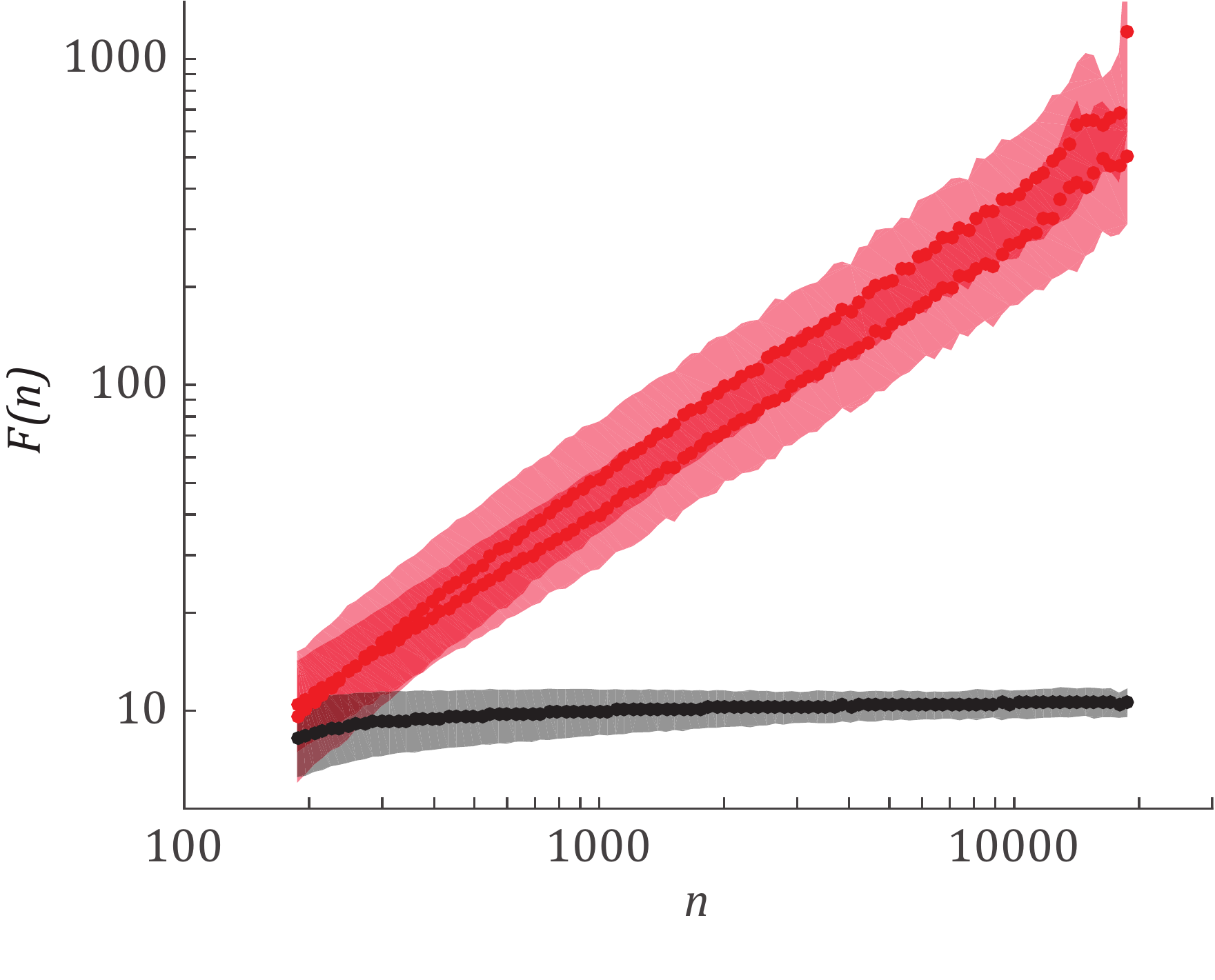} 
\end{minipage}\hfill
\begin{minipage}[c]{0.45\linewidth}
\caption{ \small {\bf Figure \ref{fig4:fig6}:} DFA mean square displacements of $Y\left( t,f \right)$ (black) and $A\left( t,f \right)$ (red; alpha and beta band, both already shown in
{\bf Figure \ref{fig4:fig3}}). As before shaded areas refer to the $25^{\text{th}}$ and $75^{\text{th}}$ percentiles of the subject averaged densities $\overline{p}_n$ } \label{fig4:fig6}
\end{minipage}
\end{figure}

To further highlight the peculiar role of amplitude and phase, we finally contrasted our results with the scaling of fluctuations of the mean MEG activity. For this, we applied our DFA to $Y\left( t,f \right) = \frac{1}{90} \sum_{k = 1}^{90} y_{k}\left( t,f \right) $, i.e. we considered not the amplitude and phase but the `raw' MEG signals. This mean activity did not display long-range correlations but rather anti-persistent ones ($\alpha = 0.02$, {\bf Figure \ref{fig4:fig6}}).
\begin{table}[h!]
\centering
\caption{ \small {\bf Table \ref{tab4:tab3}:} Relative values $\Delta \text{BIC}  = \text{BIC}  - \min( \text{BIC})$ for all variables in the beta frequency band; cf. {\bf Table \ref{tab4:tab2}}. In all cases the linear model $f_{\theta}^1$ resulted in minimal BIC values indicating power-law scaling in all variables. Corresponding $\Delta \text{AIC}_c$ values are given between brackets.} \label{tab4:tab3}
\begin{tabularx}{\textwidth}{p{0.2\textwidth} r r r r}
\toprule
$f_{\theta}^{1}\left( x \right)$ & 0.00 (0.00) & 0.00 (0.00) & 0.00 (0.00) & 0.00 (0.10) \\
$f_{\theta}^{2}\left( x \right)$ & 0.37 (0.37) & 2.72 (2.72) & 0.35 (0.35)& 1.36 (1.36) \\
$f_{\theta}^{3}\left( x \right)$ & 4.04 (1.57) & 4.60 (2.12) & 4.17 (1.69) & 4.50 (2.02) \\
$f_{\theta}^{4}\left( x \right)$ & 6.29 (6.29) & 11.46 (11.46) & 5.33 (5.33)& 6.83 (6.83) \\
$f_{\theta}^{5}\left( x \right)$ & 3.99 (1.51) & 4.60 (2.12) & 4.10 (1.62) & 4.49 (2.01) \\
$f_{\theta}^{6}\left( x \right)$ & 4.20 (1.72) & 4.57 (2.09) & 4.33 (1.85) & 4.63 (2.15) \\
$f_{\theta}^{7}\left( x \right)$ & 8.15 (3.24) & 9.18 (4.27) & 8.36 (3.44) & 8.89 (3.98) \\
$f_{\theta}^{8}\left( x \right)$ & 4.02 (1.54) & 4.62 (2.14) & 4.14 (1.66) & 4.52 (2.04) \\
$f_{\theta}^{9}\left( x \right)$ & 55.24 (52.76) & 10.34 (7.86) & 26.24 (23.76) &12.97 (10.49) \\
\bottomrule
\end{tabularx}
\end{table} \\
\vspace{\fill}
\hfill
\section{Discussion}
We report power-law scaling in both amplitude and phase of collective neural activity on long time scales, which is consistent with the hypothesis that the brain operates in a critical state. Operating in a critical state is not the only way a system can generate power-law scaling. Systems in subcritical states \cite{Botcharovaetal2014} or merely stochastic systems \cite{Touboul&Destexhe2010, Touboul&Destexhe2015} may also display power laws. Biological systems display sub- and supercritical dynamics but they can be tuned into criticality \cite{Mazzonietal2007, Shew&Plenz2013, Fagerholmetal2015}. We favor the interpretation of critical states, also because it is consistent with scale-free auto-correlation structures of single channel EEG activity \cite{LinkenkaerHansenetal2001, LinkenkaerHansenetal2007}, size and duration of neural avalanches \cite{Beggs&Plenz2003, Rubinovetal2011, Shrikietal2013} and, in the spatial domain, degree distributions of neuroanatomical and functional connectivity networks \cite{Eguiluzetal2005, Bullmore&Sporns2009}. Such scaling laws in neuronal dynamics are also correlated with those found in behavior \cite{Palvaetal2013}.

Previous studies addressing power-law scaling in neuronal dynamics were solely based on spatially local measures. For instance, power-law scaling has been reported using pair-wise synchronization measures like PLI and $\Delta^{2}\left( t, \Delta t \right)$ by, e.g., Kitzbichler and coworkers \cite{Kitzbichleretal2009, Kitzbichleretal2015} and Farmer \cite{Farmer2015}. In contrast, we evaluated scaling behavior on a global brain scale by using overall amplitude and phase synchrony as collective variables. Analyzing the scaling behavior of these variables enabled us to directly compare phase and amplitude behavior in brain activity.

What does the difference in power-law scaling of amplitudes and phases tell us about information processing in the brain? Despite the fact that amplitude and phase differ in their perseverance, both show long-range correlations over a scale of hundreds of seconds, suggesting that they reflect `memory' of cortical states. A higher scaling exponent reflects a slower decay of auto-correlations and hence a more predictable signal with decreased entropy \cite{Carbone&Stanley2007}. Therefore, we speculate that the difference in scaling exponents and their associated complexity reflect that amplitude and phase play different roles in information processing and memory decoding: (low-fidelity) amplitude dynamics decode long-term memory, whereas the (high-fidelity) phase synchrony comprises a more complex and flexible memory coding (in an information theoretic sense).

Volume conduction can be a confounder in analyzing encephalographic recordings \cite{Schoffelen&Gross2009}. Several methods to mitigate its effects have been proposed, generally relying on removing the instantaneous interactions that signify volume conduction \cite{Nolteetal2004, Stametal2007, Brookesetal2012}. In consequence these methods can only be applied in a pair-wise fashion, such that they are not applicable when considering the variables $R^{\left( \cdot \right)}\left( t,f \right)$ and $A^{\left( \cdot \right)}\left( t,f \right)$. We note, however, that volume conduction does not significantly influence the auto-correlation structure of the signals under study. This finds support by Shriki and coworkers \cite{Shrikietal2013} who showed that mere linear mixing cannot `transform' uncorrelated activity to power-law scaling. In fact, if activity displays a power law, linear mixing does not alter this apart from slightly lowering the scaling exponent.

The occurrence of power laws is not only consistent with the 'criticality hypothesis' \cite{Beggs2008}. The macroscopic behavior of self-organizing processes can | in general | be cast into a low-dimensional system when critical \cite{Haken1977}; this in fact motivated looking at collective amplitude and phase synchrony. Several modeling studies support the seminal role of self-organization in neural dynamics and often highlight self-organized criticality. For example, synaptic plasticity under the influence of a simple learning rule leads to scale-free networks \cite{Christensenetal1998, Bornholdt&Roehl2003, Drosteetal2012} and power-law distributions of avalanche dynamics \cite{Rubinovetal2011}. In resting state, the mechanism of local feedback mediated inhibition increases model performance \cite{Decoetal2014} and may also be interpreted as a form of self organization.
Furthermore, the slow evolution of the order parameters in self-organizing systems is consistent with the time scale on which RSNs evolve \cite{Biswaletal1995, Raichleetal2001, Fox&Raichle2007, Brookesetal2011}.

While previous work has shown that RSNs fluctuate at slow ($>1$sec) time scales \cite{Biswaletal1995, Brookesetal2011}, it has been recently shown that RSNs are also expressed in MEG activity on faster ($<1$sec) time scales \cite{Bakeretal2014}. Alongside the results in this paper, this represents mounting evidence that RSNs are expressed across a range of time scales, i.e. they are time scale invariant.

In summary we have shown the presence of persistent long-range correlations in the evolution of global brain dynamics, i.e. the auto-correlation function obeys a power-law with scaling exponents exceeding those corresponding to random processes without memory. This adds further support to the hypothesis that the brain is in a (permanently) critical state. The here-reported scaling exponents clearly discriminate amplitude and phase dynamics, suggesting their differential role in whole-brain information processing.



\end{document}